\documentclass{article}
\usepackage[german, english]{babel}
\usepackage[a4paper, left=2.5cm, right=2.5cm, top=3.5cm, bottom=3cm]{geometry}
\usepackage{caption}
\usepackage{mathtools}
\usepackage{wrapfig}
\newsavebox{\arrangebox}
\newlength{\arrangeht}
\usepackage{amssymb}
\usepackage{amsmath}
\usepackage{amscd}
\usepackage{latexsym}
\usepackage{graphicx}
\usepackage{ifthen}
\usepackage{epsfig}
\usepackage{setspace}
\usepackage{subcaption}
\usepackage{hyperref}
\usepackage{dsfont}
\usepackage{mathrsfs}
\usepackage{color}
\usepackage{cancel}
\usepackage{mathtools}
\usepackage{wrapfig}
\hypersetup{
 colorlinks=true,
 linkcolor=blue,
 citecolor=magenta,
}

\catcode`@=11
\renewcommand{\section}{\@startsection{section}{1}{0pt}{\medskipamount}
{\medskipamount}{\large\bf}}
\numberwithin{equation}{section}
\catcode`@=12

\newcommand{\R}{\mathds R}
\newcommand{\g}{\frak g}
\newcommand{\h}{\frak h}
\newcommand{\m}{\frak m}
\newcommand{\Acal}{{\cal A}}
\newcommand{\Fcal}{{\cal F}}
\newcommand{\Ecal}{{\cal E}}
\newcommand{\Bcal}{{\cal B}}
\newcommand{\Tcal}{{\cal T}}
\newcommand{\Scal}{{\cal S}}

\def\ep{\mathrm{e}}
\def\pa{\mbox{$\partial$}}
\def\diff{\mathrm{d}}
\def\tr{\mathrm{tr}}
\def\sfrac#1#2{{\textstyle\frac{#1}{#2}}}
\def\]{\right]}
\def\[{\left[}
\def\){\right)}
\def\({\left(}
\def\>{\rangle}
\def\<{\langle}
\def\+{\dagger}

\def\={\ =\ }
\def\und{\quad\textrm{and}\quad}
\def\with{\quad\textrm{with}\quad}
\def\for{\quad\textrm{for}\quad}

\begin{document}

\title{\bf\huge 
SO(1,3) Yang--Mills solutions on Minkowski space via cosets}
\date{~}

\author{\phantom{.}\\[12pt]
\hspace{-.5cm}
{\Large Kaushlendra Kumar\footnote{kaushal.kumar224@gmail.com}}
\\[24pt]
{Institut f\"ur Theoretische Physik \&}\\ 
{Leibniz Universit\"at Hannover} \\ 
{Appelstra{\ss}e 2, 30167 Hannover, Germany}\\
[12pt] 
} 

\clearpage
\maketitle

\begin{abstract}
\noindent\large
We present a novel family of Yang--Mills solutions, with gauge group SO(1,3), on Minkowski space that are geometrically distinguished into two classes, viz.~interior and exterior of the lightcone. We achieve this by foliating the former with SO(1,3)/SO(3) cosets and the latter with SO(1,3)/SO(1,2) cosets and analytically solving the Yang--Mills equation of an SO(1,3)-invariant gauge field. The resulting fields and their stress-energy tensor, when translated to the Minkowski space, diverge at the lightcone, but we demonstrate how this stress-energy tensor could be regularised due to its unique algebraic structure.
\end{abstract}

\medskip

\section{Introduction}
\label{sec:intro}
Analytic solutions of Yang-Mills theory with compact gauge groups, such as SU(2), and finite action are very few, like the ones presented in \cite{Luescher,AFF,Schechter} or more recently in \cite{ILP17,ILP17(2),IL08}. Here we improve upon this situation by presenting new solutions \cite{KLPR22}, albeit with a non-compact gauge group $G{=}\text{SO(1,3)}$, i.e.~the Lorentz group. The latter appears in a gauge theory formulation of general relativity and could be relevant for emergent/modified theories of gravity including supergravity and matrix models. 

The construction of these solutions relies on the fact
that, owing to the natural action of the Lorentz group on Minkowski space $\R^{1,3}$, there exists foliations of $\R^{1,3}$ into $G$-orbits that are reductive and symmetric coset spaces $G/H$. Specifically, on the inside of the lightcone we have $H{=}\text{SO(3)}$ and on the outside of the lightcone we have $H{=}\text{SO(1,2)}$. The former is the Riemannian two-sheeted hyperbolic space $H^3$, foliating the future and past of the lightcone with timelike parameter $u$, while the latter is a pseudo-Riemmanian de Sitter space $\text{dS}_3$, foliating the exterior of lightcone with spacelike parameter $u$. The Yang--Mills dynamics on these spaces are separately studied by considering a $G$-invariant gauge connection $\Acal$ and employing dimensional reduction on $\R\times G/H$ that yields a Newton's particle, parameterized by $u$ and subject to an inverted double-well potential, which admits analytic solutions.

These field configurations are then pulled back to the respective domains of the Minkowski space, yielding the color-electromagnetic fields---diverging at the lightcone---that we then use to compute the stress-energy tensor. The latter turns out to have the same form in both the domains; this form, curiously, takes the shape of a pure {\it improvement term}. This fact can be used to regularize the stress-energy tensor across the lightcone so that one can have matching of the fields, defined on two domains of the spacetime, at the lightcone.

\section{Minkowskian geometry and its foliations}
\begin{figure}[h]
\centering

\sbox{\arrangebox}{%
  \begin{subfigure}[b]{0.45\columnwidth}
  \centering
  \includegraphics[width=.9\textwidth,height=10cm]{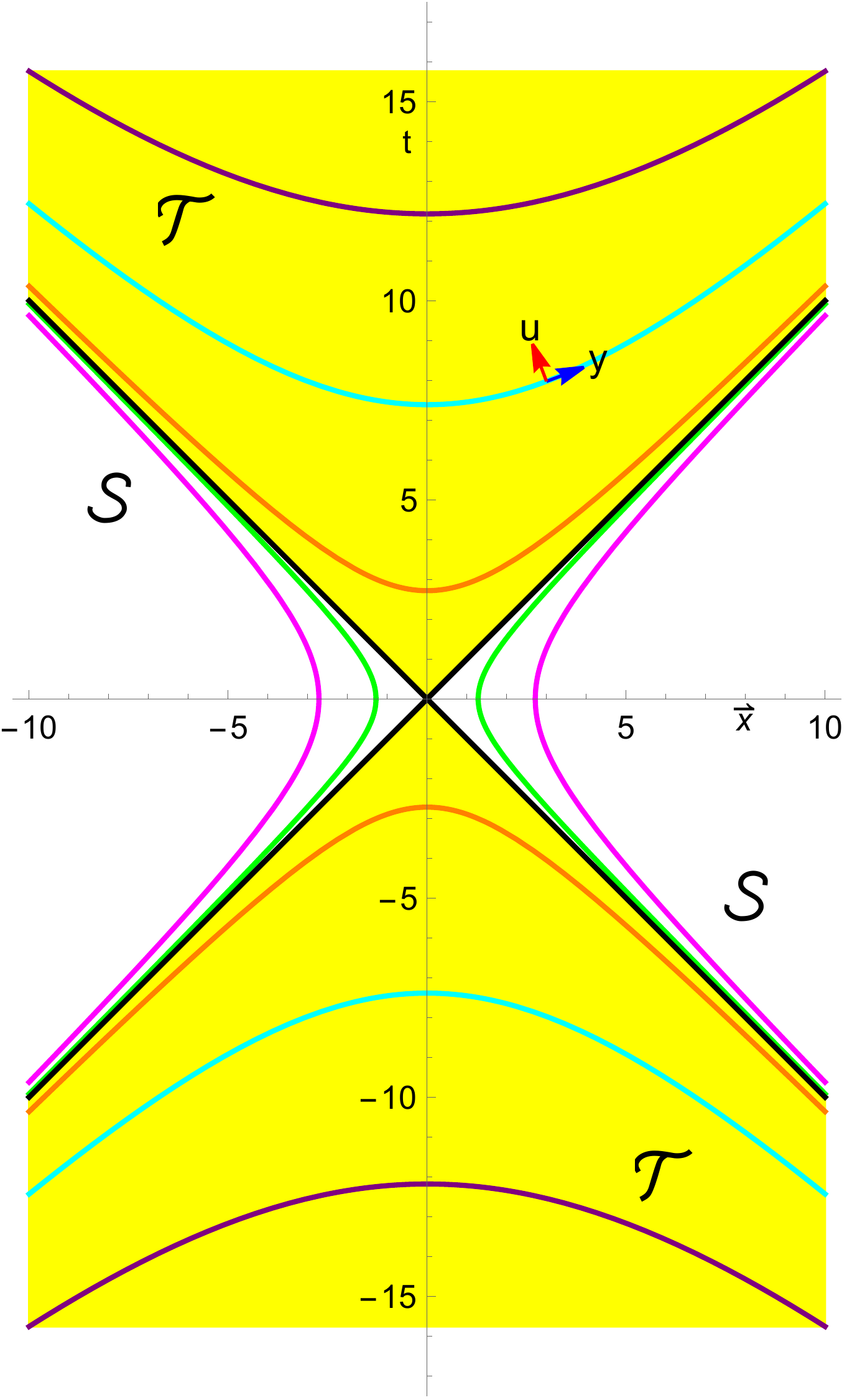}
  \caption{Foliation of interior (yellow region) and exterior of the lightcone with $H^3$- and $\text{dS}_3$-slices respectively. Every slicing has internal coordinate $y$ and foliation parameter $u$.}
  \label{MinkFol}
  \end{subfigure}%
}
\setlength{\arrangeht}{\ht\arrangebox}

\usebox{\arrangebox}\hfill
\begin{minipage}[b][\arrangeht][s]{0.45\columnwidth}
  \begin{subfigure}[t]{\textwidth}
  \centering
  \vspace{-10pt}
  \includegraphics[width=.9\textwidth,height=4cm]{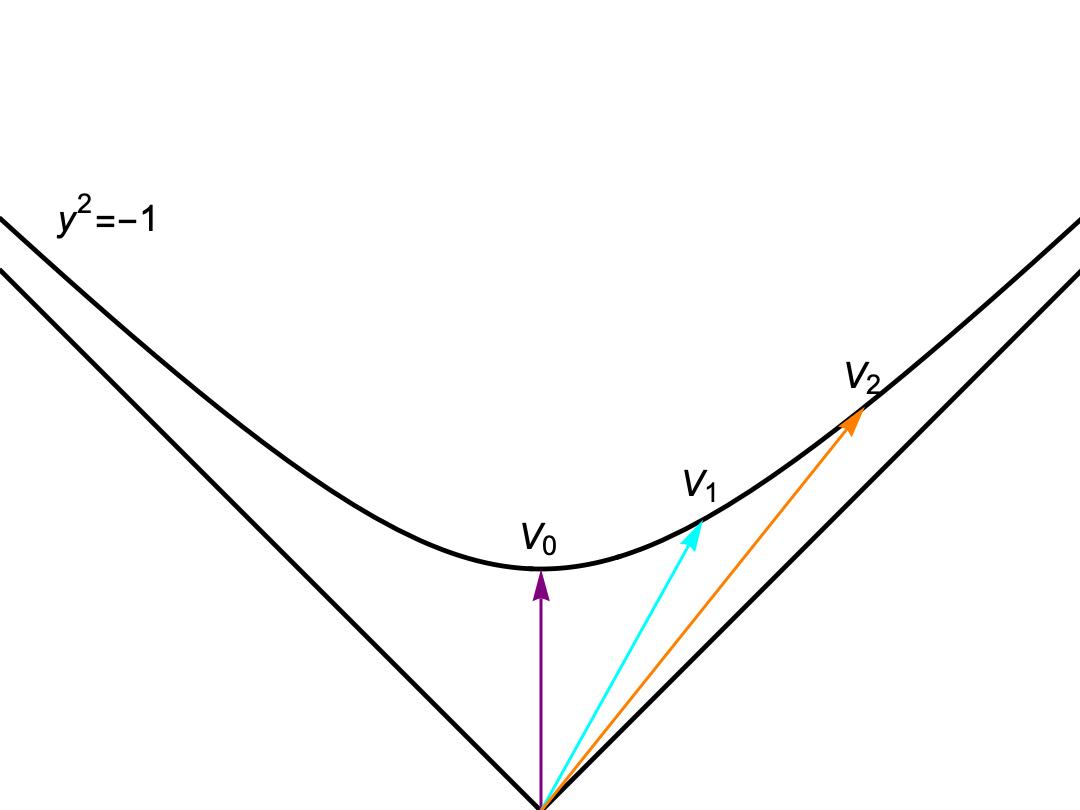}
  \caption{Every $H^3$ vector $V_\alpha$ is related to $V_0\sim (1,0,0,0)^\top$ with a unique boost $\Lambda_\alpha$, yielding its stability subgroup: $\Lambda_\alpha\,\mathrm{SO}(3)\,\Lambda_\alpha^{\text -1}$.}
  \label{H3vectors}
  \end{subfigure}\vfill
  \begin{subfigure}[b]{\textwidth}
  \centering
  \includegraphics[width=.9\textwidth,height=4cm]{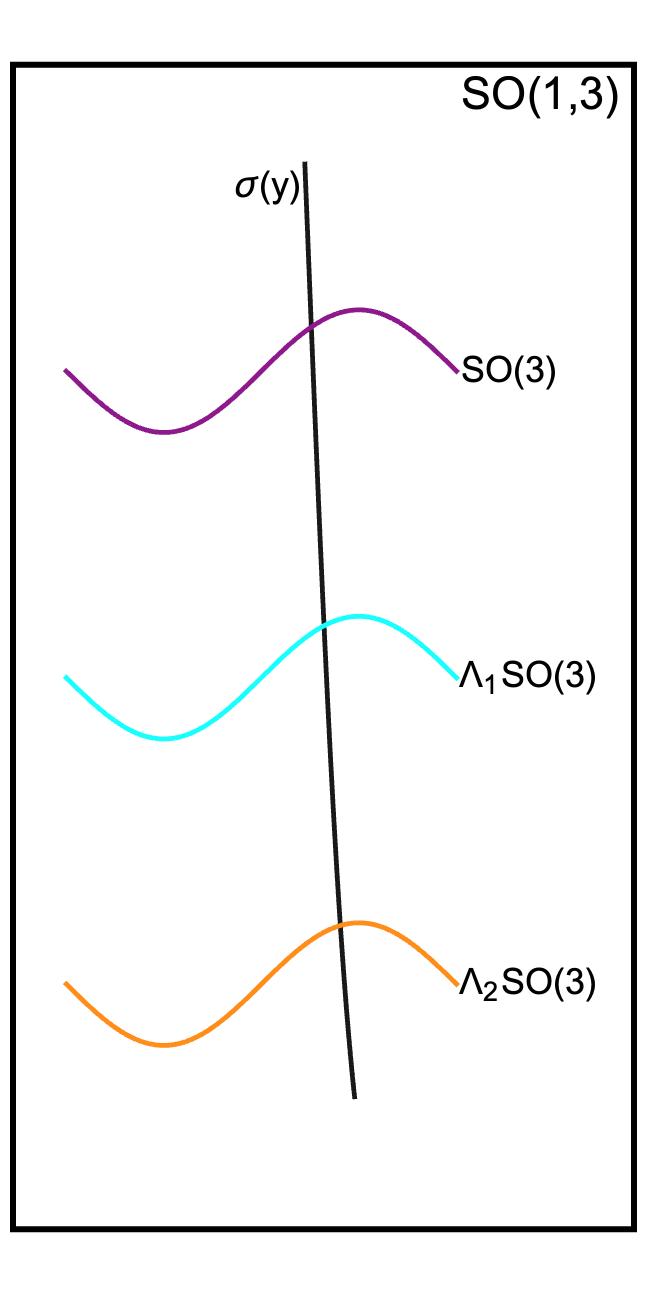}
  \caption{The above $H^3$ vectors $V_\alpha$ lies in one-to-one correspondence with cosets $\Lambda_\alpha\,\mathrm{SO}(3)$ so that $H^3$, as a 3-dimensional submanifold $\sigma(y)$ inside SO(1,3), arises through a choice of representative~$\sigma$ in each of these cosets.}
  \label{CosetspaceH3}
  \end{subfigure}
\end{minipage}
\caption{Minkowski foliations and demonstration of $H^3\cong \textrm{SO(1,3)/SO(3)}$.}
\end{figure}

We can foliate the Minkowski space $\R^{1,3}$, with metric $(\eta_{\mu\nu}) = (-,+,+,+)$ for $\mu,\nu=0,1,2,3$, in two parts as depicted in Figure \ref{MinkFol}: (a) the lightcone-interior $\Tcal$ with two-sheeted hyperbolic space $H^3$ and (b) the lightcone-exterior $\Scal$ with single-sheeted de Sitter space $\text{dS}_3$. 

For the first case, the hyperbolic space $H^3$ is embedded in $\R^{1,3}$ algebraically as
\begin{equation} \label{H3}
    y{\cdot}y\equiv \eta_{\mu\nu}\,y^{\mu}\,y^{\nu} \= -1\ ,
\end{equation}
and foliates---with a timelike parameter $u$ obeying $\ep^u = \sqrt{|x{\cdot}x|}$---the lightcone-interior $\Tcal$ as\footnote{We employ standard conventions $x^0{=}t, x^1{=}x, x^2{=}y, x^3{=}z$, $\vec{x} := (x^1,x^2,x^3)$ and $r = \sqrt{\vec{x}{\cdot}\vec{x}}$ in this article.}
\begin{equation}\label{Tfolitation}
\begin{aligned}
    \varphi_{_\Tcal} &:~ \mathds{R}\times H^3 \rightarrow \Tcal\ , \quad (u,y^\mu) \mapsto x^\mu := \ep^u\,y^\mu\ , \\
    \varphi_{_\Tcal}^{-1} &:~ \Tcal \rightarrow \mathds{R}\times H^3\ ,\quad x^\mu \mapsto (u,y^{\mu}) := \Bigl(\ln{\sqrt{|x{\cdot}x|}},\frac{x^\mu}{\sqrt{|x{\cdot}x|}} \Bigr)\ .
\end{aligned}
\end{equation}
With this, the metric on $\Tcal$ becomes conformal to a Lorentzian cylinder $\mathds{R}\times H^3$:
\begin{equation}\label{metricT}
    \diff{s}_{_\Tcal}^2 \= \ep^{2u}\left( -\diff{u}^2 + \diff{s}_{H^3}^2 \right)\ ,
\end{equation}
where $\diff{s}_{H^3}^2$ is the flat metric on $H^3$ arising from \eqref{H3}. 

In the second case, we can embed $\text{dS}_3$ inside the Minkowski space $\R^{1,3}$ by
\begin{equation}\label{dS3}
    y\cdot y \equiv \eta_{\mu\nu}\,y^\mu\,y^\nu \= 1\ . 
\end{equation}
The foliation of $\Scal$ follows analogous to the previous case, albeit with a spacelike foliation parameter $u$ satisfying $\ep^u = \sqrt{|x{\cdot}x|}\equiv \sqrt{r^2-t^2}$:
\begin{equation}\label{Sfolitation}
\begin{aligned}
    \varphi_{_\Scal} &:~ \R\times\text{dS}_3 \rightarrow \Scal\ , \quad (u,y^\mu) \mapsto x^\mu := \ep^u\,y^\mu \ , \\
    \varphi_{_\Scal}^{-1} &:~ \Scal \rightarrow \R\times\text{dS}_3\ ,\quad x^\mu \mapsto (u,y^\mu) := \Bigl(\ln{\sqrt{|x{\cdot}x|}},\frac{x^\mu}{\sqrt{|x{\cdot}x|}} \Bigr)\ ,
    \end{aligned}
\end{equation}
such that the metric on $\Scal$ becomes conformal to the metric on a cylinder $\R\times\text{dS}_3$,
\begin{equation}\label{metricS}
    \diff{s}_{_\Scal}^2 \= \ep^{2u}\left( \diff{u}^2 + \diff{s}_{\text{dS}_3}^2 \right)\ ,
\end{equation}
where the flat $\text{dS}_3$-metric $\diff{s}_{\text{dS}_3}^2$ is induced from \eqref{dS3}.

\section{Algebra/geometry of symmetric SO(1,3)-cosets}
We consider here reductive coset spaces $G/H$ with 6-dimensional Lie group $G=\textrm{SO(1,3)}$ that are also symmetric. This yields homogeneous spaces $H^3$ and $\text{dS}_3$ with stablity subgroups $H{=}\textrm{SO(3)}$ and $H{=}SO(1,2)$ respectively. The equivalence of SO(1,3)/SO(3) with $H^3$ is geometrically illustrated through Figures \ref{H3vectors} and \ref{CosetspaceH3}. 

The six generators $\lbrace I_{\scriptscriptstyle{A}}\rbrace$ of the Lie algebra $\g=\text{Lie}(G)$ are nothing but the canonical rotation ($J_a$) and boost ($K_a$) generators of the Lorentz group:
\begin{equation}
    J_1 {=} \begin{psmallmatrix} 0 & 0 & 0 & 0 \\ 0 & 0 & 0 & 0 \\ 0 & 0 & 0 & {\text -1} \\ 0 & 0 & 1 & 0  \end{psmallmatrix}, J_2 {=} \begin{psmallmatrix} 0 & 0 & 0 & 0 \\ 0 & 0 & 0 & 1 \\ 0 & 0 & 0 & 0 \\ 0 & {\text -1} & 0 & 0  \end{psmallmatrix}, J_3 {=} \begin{psmallmatrix} 0 & 0 & 0 & 0 \\ 0 & 0 & {\text -1} & 0 \\ 0 & 1 & 0 & 0 \\ 0 & 0 & 0 & 0  \end{psmallmatrix},
    K_1 {=} \begin{psmallmatrix} 0 & 1 & 0 & 0 \\ 1 & 0 & 0 & 0 \\ 0 & 0 & 0 & 0 \\ 0 & 0 & 0 & 0  \end{psmallmatrix}, K_2 {=} \begin{psmallmatrix} 0 & 0 & 1 & 0 \\ 0 & 0 & 0 & 0 \\ 1 & 0 & 0 & 0 \\ 0 & 0 & 0 & 0  \end{psmallmatrix}, K_3 {=} \begin{psmallmatrix} 0 & 0 & 0 & 1 \\ 0 & 0 & 0 & 0 \\ 0 & 0 & 0 & 0 \\ 1 & 0 & 0 & 0  \end{psmallmatrix}\ .
\end{equation}
Moreover, for reductive cosets, $\g$ splits into a Lie subalgebra $\h=\text{Lie}(H)$ and its orthogonal complement\footnote{The orthogonality here is with respect to the Cartan--Killing metric, given by the trace of adjoint representation of these generators $\lbrace I_{\scriptscriptstyle{A}}\rbrace$.} $\m$; this is also reflected in the splitting of the generators $\lbrace I_{\scriptscriptstyle{A}}\rbrace$ as follows
\begin{equation} \label{split}
    \g \= \h \oplus \m\quad \implies\quad \lbrace I_{\scriptscriptstyle{A}}\rbrace \= \lbrace I_{i}\rbrace \cup \lbrace I_{a}\rbrace \quad\with i = 4,5,6 \und a = 1,2,3\ ,
\end{equation}
where $\{I_i\}$ spans $\h$ and $\{I_a\}$ spans $\m$. They satisfy following commutation relations
\begin{equation} \label{LieAlgSplit}
    [I_i,I_j]\=f_{ij}^{\ \ k}I_k\ , \qquad [I_i,I_a]\=f_{ia}^{\ \ b}I_b\ , \,\und\, [I_a,I_b]\=f_{ab}^{\ \ i}I_i\ .
\end{equation}
Similarly, the left-invariant one-forms of SO(1,3) split into $\lbrace e^{\scriptscriptstyle A} \rbrace = \lbrace e^i \rbrace \cup \lbrace e^a \rbrace$, such that the following structure equations---with same structure coefficients as in \eqref{LieAlgSplit}---are satisfied
\begin{equation} \label{StrucEqns}
    \diff e^a + f_{ib}^{\ \ a}\,e^i\wedge e^b \= 0 \qquad\und\qquad \diff e^i + \sfrac12 f_{jk}^{\ \ i}\,e^j\wedge e^k + \sfrac12 f_{ab}^{\ \ i}\,e^a\wedge e^b \= 0\ .
\end{equation}
Here $e^a$ yields the metric on $G/H$ while $e^i = e^i_a\,e^a$ are linearly dependent. 

\subsection{Case I: $H^3\cong \textrm{SO(1,3)/SO(3)}$}
For SO(1,3)/SO(3) the splitting \eqref{split} and structure coefficients \eqref{LieAlgSplit} are given by
\begin{equation}
   I_i = J_{i-3}\ ,\ I_a = K_a \quad \implies\quad f_{ij}^{\ \ k} \= \varepsilon_{i-3\;j-3\;k-3}\ ,\ f_{ia}^{\ \ b} \= \varepsilon_{i-3\;a\,b}\ ,\    f_{ab}^{\ \ i} \= -\varepsilon_{a\,b\,i-3}\ .
\end{equation}
The identification of coset space SO(1,3)/SO(3) with $H^3$ is seen from the following maps
\begin{equation}
\begin{aligned}
    \alpha_{_\Tcal} &:~ \mathrm{SO}(1,3)/\mathrm{SO}(3) \rightarrow H^3\ ,\quad [\Lambda_\Tcal] \mapsto y^\mu = (\Lambda_{\Tcal})^{\mu}_{\;\;0}\ ,\\ 
    \alpha_{_\Tcal}^{-1} &:~ H^3 \rightarrow \mathrm{SO}(1,3)/\mathrm{SO}(3)\ , \quad y^\mu \mapsto [\Lambda_\Tcal]\ ,
\end{aligned}    
\end{equation}
where the representative $\Lambda_\Tcal$ of the coset $[\Lambda_\Tcal]:=\{\Lambda = \Lambda_\Tcal h: h\in SO(3)\}$ is given by
\begin{equation}\label{matrixT}
\Lambda_\Tcal \= \begin{pmatrix} \gamma & \gamma\,\pmb{\beta}^\top \\ \gamma\,\pmb{\beta} & \mathds{1} + (\gamma{-}1)\frac{\pmb{\beta}\otimes\pmb{\beta}}{\pmb{\beta}^2} \end{pmatrix}\ , \quad\with
\beta^a = \frac{y^a}{y^0}\ ,\quad \gamma = \frac{1}{\sqrt{1-\pmb{\beta}^2}} = y^0\ ,
\end{equation}
and $\pmb{\beta}^2=\delta_{ab}\,\beta^a\beta^b\ge0$. It is straightforward to verify that the above map $\alpha_{_\Tcal}$ is well-defined and, in fact, $\Lambda_\Tcal$ is a generic boost obtained from coset generators $I_a\in \m$ as
\begin{equation}\label{Texp}
    \Lambda_\Tcal \= \textrm{exp}(\eta^a\,I_a)\ , \quad\with \beta^a = \sfrac{\eta^a}{\sqrt{\pmb{\eta}^2}}\tanh{\sqrt{\pmb{\eta}^2}}\ , \for \pmb{\eta}^2 = \delta_{ab}\,\eta^a\eta^b\ .
\end{equation}
We can now obtain the left-invariant one-forms by employing the Maurer--Cartan prescription:
\begin{equation}
    \Lambda_\Tcal^{-1}\,\diff{\Lambda}_\Tcal \= e^a\,I_a + e^i\,I_i\ ;\quad e^a = \Bigl(\delta^{ab} - \frac{y^a\,y^b}{y^0(1{+}y^0)}\Bigr)\,\mathrm{d}y^b \und e^i = \varepsilon_{i-3\;a\,b}\,\frac{y^a}{1{+}y^0}\,\mathrm{d}y^b\ ,
\end{equation}
such that $e^a$ reproduces the metric on $H^3$ while $e^i$ become linearly dependent as follows
\begin{equation}\label{1formRel1}
    \diff{s}_{H^3}^2 \= \delta_{ab}\,e^a\otimes e^b \,\und\, e^i = e^i_a\,e^a\ , \with e^i_a = \varepsilon_{a\,i-3\;b}\,\frac{y^b}{1{+}y^0}\ .
\end{equation}

\subsection{Case II: $\text{dS}_3\cong$ SO(1,3)/SO(1,2)}
For the coset space $\mathrm{SO}(1,3)/\mathrm{SO}(1,2)$ we chose the splitting \eqref{split} as follows
\begin{equation}
    I_i \in \lbrace K_1, K_2, J_3 \rbrace
    \,\und\,
    I_a \in \lbrace J_1, J_2, K_3 \rbrace \ ,
\end{equation}
such that the structure coefficients \eqref{LieAlgSplit} comes out to be
\begin{equation}
    f_{ij}^{\ \ k} \= \varepsilon_{i-3\;j-3\;k-3}\,(1{-}2\,\delta_{k6})\ ,\qquad f_{ia}^{\ \ b} \= \varepsilon_{i-3\;a\,b}\,(1{-}2\,\delta_{a3}) \,\und\,  f_{ab}^{\ \ i} \= \varepsilon_{a\,b\,i-3}\ ,
\end{equation}
where no summation convention is used inside the brackets. As before, we demonstrate the equivalence between $\text{dS}_3$ and $\mathrm{SO}(1,3)/\mathrm{SO}(1,2)$ through following well-defined maps
\begin{equation}
\begin{aligned}
    \alpha_{_\Scal} &:~ \mathrm{SO}(1,3)/\mathrm{SO}(1,2) \rightarrow \diff{\text{S}}_3\ ,\quad [\Lambda_\Scal] \mapsto y^\mu := (\Lambda_\Scal)^{\mu}_{\;\;3}\ , \\
    \alpha_{_\Scal}^{-1} &:~ \diff{\text{S}}_3 \rightarrow \mathrm{SO}(1,3)/\mathrm{SO}(1,2)\ , \quad y^\mu \mapsto [\Lambda_\Scal]\ ,
\end{aligned}    
\end{equation}
where the representative left-coset element $\Lambda_\Scal$ is again obtained though exponentiation with coset generators $\lbrace J_1, J_2, K_3 \rbrace$. The resultant Maurer--Cartan one-forms look like
\begin{equation}\label{1formS}
    \Lambda_\Scal^{-1}\,\diff{\Lambda}_\Scal \= e^a\,I_a + e^i\,I_i\ ;\quad e^a = \diff{y}^{3-a} - \frac{y^{3-a}}{1{+}y^3}\,\diff{y}^3 \und e^i = -\varepsilon_{i-3\;a\,b}\,\frac{y^{3-a}}{1{+}y^3}\,\diff{y}^{3-b}\ .
\end{equation}
These one-forms behave as expected with $(\eta_{ab}){=}(-,+,+)$ due to the stabilizer SO(1,2):
\begin{equation}\label{1formRel2}
    \diff{s}_{\diff{\text{S}}_3}^2 \= \eta_{ab}\,e^a\otimes e^b \,\und\, e^i = e^i_a\,e^a\ , \with e^i_a = \varepsilon_{i-3\;a\,b}\,\frac{y^{3-b}}{1{+}y^3}\ .
\end{equation}

\subsection{The lightcone exception}
Before we move on, let us make a remark on the lightcone itself that sits here as an exception in the following manner. It can be shown that both future and past of the lightcone are individually isomorphic to SO(1,3)/ISO(2), where the stability subgroup is a Euclidean group $\mathrm{E}(2) {=} \mathrm{ISO}(2)$ generated by two translations and one rotation. However, this coset space is non-reductive and, on top of that, this does not give rise to any foliation here, making this case unsuitable to study Yang--Mills dynamics as discussed in Section \ref{secYMdynamics}.

\section{Yang--Mills fields from dimensional reduction}
\label{secYMdynamics}
The study of Yang--Mills dynamics on $\R\times G/H$ via dimensional reduction is a well-known topic with excellent review in \cite{KZ92}. Given an orthonormal frame $\lbrace e^u{:=} \diff{u}, e^a \rbrace$ on the cylinder $\R\times G/H$, we can write a generic connection one-form $\Acal$ in the ``temporal" gauge $\Acal_u{=}0$ and its curvature two-form $\Fcal = \diff{\Acal} + \Acal\wedge\Acal$ as follows
\begin{equation}
    \Acal \= \Acal_a\,e^a \qquad\implies\qquad \Fcal = \Fcal_{ua}\,e^u\wedge e^a + \sfrac12 \Fcal_{ab}\,e^a\wedge e^b\ .
\end{equation}
Next, we expand the gauge field $\Acal_a$ in terms of full SO(1,3)-generators \eqref{split} as $\Acal_a = \Acal_a^i\,I_i + \Acal_a^b\,I_b$ and impose $G$-invariance on this, yielding following two conditions~\footnote{ We can write the second relation more succinctly as $[I_i,\widetilde{\Acal}_a] = f_{ia}^{\ \ b}\widetilde{\Acal}_b$ for $\widetilde{\Acal}_a := \Acal_b^a\,I_a \in \m$.}
\begin{equation} \label{G-equiv}
    \Acal_a^i \= e^i_a \,\und\, \Acal_b^a=\Acal_b^a(u)\ , \with f_{ia}^{\ \ c}\Acal_b^a \= f_{ib}^{\ \ a}\Acal_a^c\ .
\end{equation}
Furthermore, for the symmetric spaces that concerns us, one finds that $\Acal^a_b(u) = \phi(u)\,\delta^a_b$ such that our $G$-invariant gauge field $\Acal$ depends on a single real function~$\phi$:
\begin{equation} \label{ansatz}
    \Acal \= I_i\,e^i + \phi(u)\,I_a\,e^a\ .
\end{equation}
The components of the field strength $\Fcal$, using \eqref{G-equiv} and \eqref{StrucEqns}, computes to
\begin{equation} \label{Fcal}
    \Fcal_{ua} \= \dot{\phi}\,I_a  \,\und\, \Fcal_{ab} \= (\phi^2{-}1)f_{ab}^{\ \ i}\,I_i\ , \quad\with \dot{\phi}:=\partial_u \phi\ ,
\end{equation}
yielding the the color-electric field $\Ecal_a = \Fcal_{au} \in \m$ and -magnetic field $\Bcal_a = \sfrac12 \varepsilon_{abc}\Fcal_{bc} \in \h$ on the cylinder. Finally, to work out the dynamics of $\phi(u)$ we look at the Yang--Mills action
\begin{equation}
    S_{\text{YM}} \= -\frac{1}{4g^2}\int \tr_\text{ad}{\left(\Fcal\wedge*\Fcal\right)}\ ,
\end{equation}
which simplifies drastically in both cases, viz.~interior of the lightcone $\Tcal$ with $M^3:=H^3$ and exterior of the lightcone with $M^3:=\text{dS}_3$, as follows
\begin{equation}\label{YMaction}
S_{\text{YM}} \= \frac{6}{g^2}\int_{\mathds{R}\times M^3}\!\!\!\diff{\text{vol}}\ \bigl(\tfrac{1}{2}\dot{\phi}^2-V(\phi)\bigr)\ ;\qquad V(\phi) \= -\sfrac12 (\phi^2{-}1)^2
\end{equation}
\begin{wrapfigure}[8]{r}[1pt]{5cm}
\vspace{-20pt}
\centering
\includegraphics[width=4.5cm]{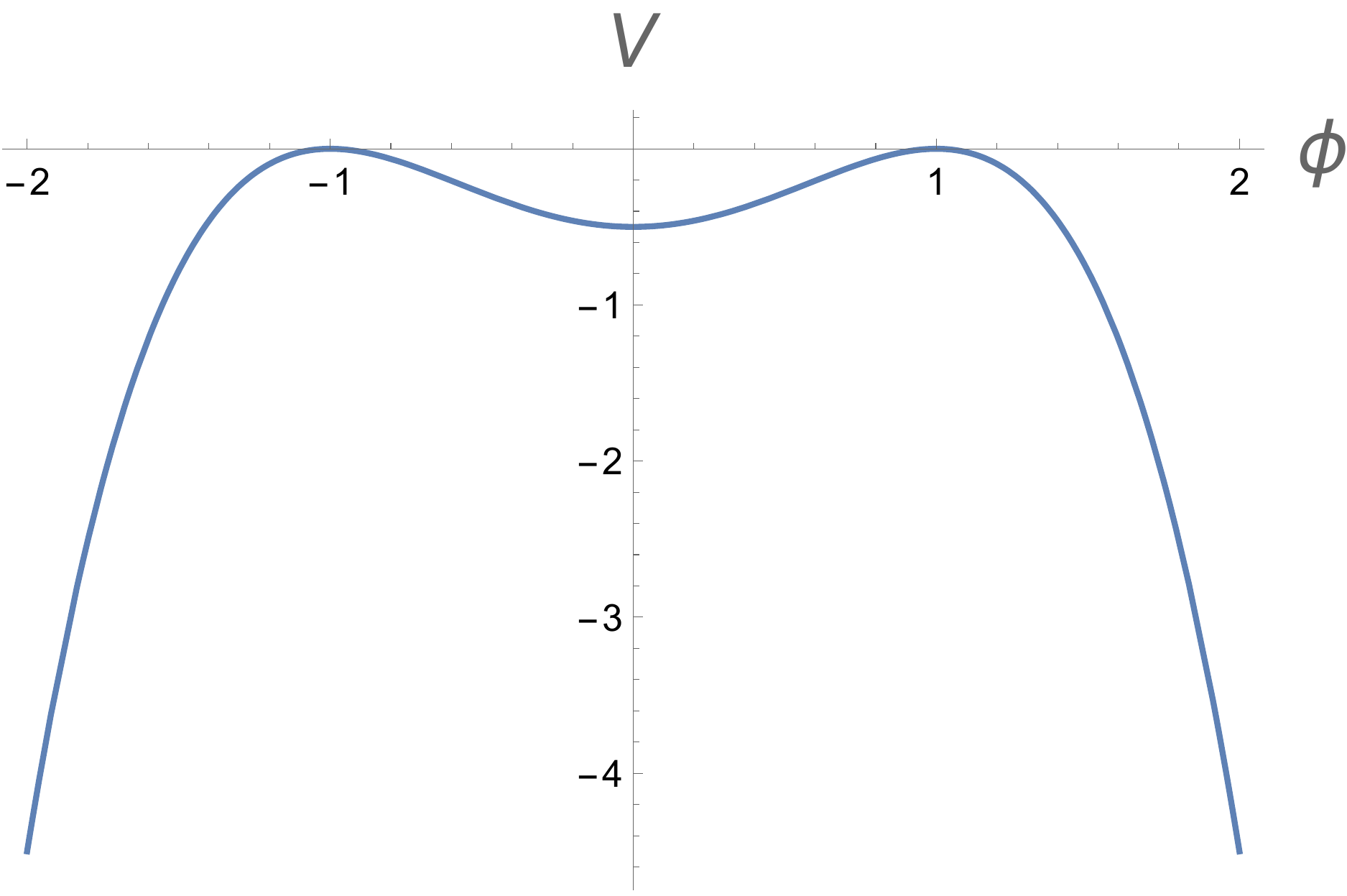}
\vspace{-5pt}
\caption{Plot of $V(\phi)$}
\label{invertedPotential}
\end{wrapfigure}
where $\diff{\text{vol}}=\sfrac{1}{3!}\varepsilon_{abc}\,\diff{u}\wedge e^a\wedge e^b\wedge e^c$ is the volume form. We immediately observe that the above action represents a mechanical particle $\phi(u)$ in an {\em inverted\/} double-well potential $V(\phi)$ depicted in Figure \ref{invertedPotential}, which yields the equation of motion for an anharmonic oscillator,
\begin{equation}\label{Teom}
    \ddot{\phi} \= - \frac{\partial V}{\partial \phi} \= 2\,\phi\,(\phi^2{-}1) \ .
\end{equation}
This admits analytic solutions in terms of Jacobi elliptic functions. For example, in the bounded case where the mechanical energy $\sfrac12 \dot{\phi}^2 + V(\phi)=:\epsilon\in[-\frac12,0]$ we have
\begin{equation} \label{YMsoln}
    \phi_{\epsilon,u_0}(u) \= f_{-}(\epsilon)\,\text{sn}\big(f_{+}(\epsilon)(u{-}u_0),k\big) \with f_{\pm}(\epsilon) = \sqrt{1\pm\sqrt{-2\epsilon}}\ ,\quad k^2 = \frac{f_{-}(\epsilon)}{f_{+}(\epsilon)}
\end{equation}
and a `time'-shift parameter $u_0$. This also include special cases, such as a ``kink'':
\begin{equation}\label{kinksoln}
    \phi \= \begin{cases}
                 \ 0 \quad &\for\epsilon = -\sfrac12\ , \\
                 \ \tanh{(u{-}u_0)}\quad &\for\epsilon = 0\ ,\\
                 \ \pm 1 \quad &\for\epsilon = 0\ .\ 
            \end{cases}
\end{equation}

Now in order to pull these solution back to $\Tcal$ we transform the orthonormal frame $\lbrace e^u, e^a \rbrace$ on $\R\times H^3$, using the map $\varphi_{_\Tcal}$ \eqref{Tfolitation} and abbreviation $|x| := \sqrt{|x{\cdot}x|}$, as follows
\begin{equation}\label{Tframe}
    e^u := \diff{u} \= \frac{t\,\diff{t} - r\,\diff{r}} {t^2 - r^2} \qquad\und\qquad 
    e^a \= \frac{1}{|x|}\Bigl( \diff{x}^a - \frac{x^a}{|x|}\,\diff{t} + \frac{x^a}{|x|(|x| + t)}\,r\,\diff{r}\Bigr)\ .
\end{equation}
The SO(1,3)-invariant gauge field $\Acal \equiv A$ \eqref{ansatz} can be casted into a Minkowski one-form
\begin{equation}\label{AcalT}
    A \= \frac{1}{|x|}\left\{ \frac{\varepsilon_{ab}^{\ \ k-3}\,x^a}{|x|+t}\diff{x}^b\,I_k + \phi(x)\Bigl( \diff{x}^a - \frac{x^a}{|x|}\,\diff{t} + \frac{x^a}{|x|(|x| + t)}\,r\,\diff{r}\Bigr)I_a  \right\}\ ,
\end{equation}
where $\phi(x) {:=} \phi_{\epsilon,u_0}(u(x))$. We can then find the field strength $F=F_{\mu\nu}\,\mathrm{d}x^\mu\wedge\mathrm{d}x^\nu$ on $\Tcal$ from its cylinder version $\Fcal$ \eqref{Fcal} using vierbein components $e^u = e^u_\mu\, \diff{x}^\mu$ and $e^a = e^a_\mu\,\diff{x}^\mu$ \eqref{Tframe}. The corresponding color-electric $E_i:=F_{0i}$ and -magnetic $B_i:=\frac{1}{2}\varepsilon_{ijk}\,F_{jk}$ fields read
\begin{equation}\label{FcalT}
\begin{aligned}
    E_a &\= \frac{1}{|x|^3}\left\{\left(\phi^2{-}1\right)\varepsilon_{ab}^{\ \ i-3}\,x^b\,I_i -\Dot{\phi} \Bigl(t\,\delta^{ab} - \frac{x^a\,x^b}{|x|+t} \Bigr)I_b \right\}\ , \\
    B_a &\= -\frac{1}{|x|^3}\left\{\left(\phi^2{-}1\right)\Bigl(t\,\delta^{a\,i-3} - \frac{x^a\,x^{i-3}}{|x|+t}\Bigr)I_i +\Dot{\phi}\, \varepsilon_{ab}^{\ \ c}\,x^b\,I_c \right\}\ .
\end{aligned}
\end{equation}
An interesting feature of these fields is the presence of color-electromagnetic duality, i.e.~$E_a \rightarrow B_a$ and $B_a \rightarrow -E_a$, which works when we simultaneously interchange $\Dot\phi\leftrightarrow(\phi^2{-}1)$ and and switch the generators as follows: $I_i \rightarrow I_a$ and $I_a \rightarrow -I_i$ (plus some obvious index adjustment). More importantly, we notice that the gauge field $\Acal$ \eqref{AcalT} along with the electric $E_i$ and magnetic $B_i$ fields \eqref{FcalT} become singular at the lightcone~$t{=}{\pm}r$. One can find out the fields on $\Scal$ using \eqref{Sfolitation} and following the same recipe as above. We restrain from reproducing these results here owing to space constraint and refer the reader to \cite{KLPR22} for explicit form of such fields.

\section{The stress-energy tensor}
We can compute the stress-energy tensor, given by the expression
\begin{equation} \label{SEtensor}
    T_{\mu\nu} \= -\tfrac{1}{2g^2}\, \tr_\text{ad}{\left( F_{\mu\alpha}\,F_{\nu\beta}\,\eta^{\alpha\beta} - \sfrac14 \eta_{\mu\nu} F^2 \right)} \,\with F^2 = F_{\mu\nu}\,F^{\mu\nu}\ ,
\end{equation}
for such Yang--Mills fields in a straightforward manner. Interestingly, we find that the computation yields the same form of stress-energy tensor on both sides of the lightcone that reads
\begin{equation}\label{stressEnergy}
    T \= \frac{\epsilon}{g^2(r^2{-}t^2)^3} \begin{pmatrix} 3t^2{+}r^2 & -4tx & -4ty & -4tz \\ -4tx & t^2{+}4x^2{-}r^2 & 4xy & 4xz \\ -4ty  & 4xy & t^2{+}4y^2{-}r^2 & 4yz \\ -4tz & 4xz & 4yz & t^2{+}4z^2{-}r^2 \end{pmatrix}\ .
\end{equation}
It is worth emphasising here that the explicit form of $\phi$, like in \eqref{YMsoln}, is irrelevant here as $T$ only depends on the total mechanical energy $\epsilon$. Moreover, this has a vanishing trace and presence of lightcone singularity, as expected. Surprisingly, this admits a nice compact form that can be recasted into a pure ``improvement'' term as follows,
\begin{equation} \label{TandS}
    T_{\mu\nu} \= \pa^\rho S_{\rho\mu\nu} \qquad\with\qquad
    S_{\rho\mu\nu} \= \frac{\epsilon}{g^2}\,
\frac{ x_\rho \eta_{\mu\nu} - x_\mu \eta_{\rho\nu} }{(x{\cdot}x)^2}\ ,
\end{equation}
where the term $S_{\rho\mu\nu}$ can be expressed using abbreviation $(\tilde{S}_\rho)_{\mu\nu} := \sfrac{g^2(x{\cdot}x)^2}{\epsilon} S_{\rho\mu\nu}$ as
\begin{equation}
    \tilde{S}_0 \= \begin{psmallmatrix} 
    0 & 0 & 0 & 0 \\
    x & -t & 0 & 0 \\
    y & 0 & -t & 0 \\
    z & 0 & 0 & -t
    \end{psmallmatrix}\ ,\quad \tilde{S}_1 \= \begin{psmallmatrix} 
    -x & t & 0 & 0 \\
    0 & 0 & 0 & 0 \\
    0 & -y & x & 0 \\
    0 & -z & 0 & x
    \end{psmallmatrix}\ ,\quad \tilde{S}_2 \= \begin{psmallmatrix} 
    -y & 0 & t & 0 \\
    0 & y & -x & 0 \\
    0 & 0 & 0 & 0 \\
    0 & 0 & -z & y
    \end{psmallmatrix}\ ,\quad \tilde{S}_3 \= \begin{psmallmatrix} 
    -z & 0 & 0 & t \\
    0 & z & 0 & -x \\
    0 & 0 & z & -y \\
    0 & 0 & 0 & 0
    \end{psmallmatrix}\ .
\end{equation}
Naturally, one hopes to glue the two expressions for stress-energy tensors to find a single expression valid across the Minkowski spacetime. The price to pay here is the singularity at the lightcone, which can be remedied through the following regularization procedure
\begin{equation} \label{EMreg}
    S^{\textrm{reg}}_{\rho\mu\nu} \= \frac{\epsilon}{g^2}\,
\frac{ x_\rho \eta_{\mu\nu} - x_\mu \eta_{\rho\nu} }{(x{\cdot}x+\delta)^2}
\qquad\Rightarrow\qquad
T^{\textrm{reg}}_{\mu\nu} \= \frac{\epsilon}{g^2}\,
\frac{4\,x_\mu x_\nu - \eta_{\mu\nu}x{\cdot}x + 3\,\delta\,\eta_{\mu\nu}}{(x{\cdot}x+\delta)^3}\ .
\end{equation}
This nonsingular improvement term (with a finite regularization parameter~$\delta$) yields vanishing energy and momenta as their fall-off behaviour at spatial infinity is fast enough. 

An alternate route to regularization could be to directly shift only the denominator of $T_{\mu\nu}$ in \eqref{TandS} via $x{\cdot}x\mapsto x{\cdot}x+\delta$. We can then improve the resultant stress-energy tensor up to the term in \eqref{EMreg} so as to obtain the following energy-momentum tensor candidate that is regular and that also vanishes as $\delta\to0$,
\begin{equation}
    T^\delta_{\mu\nu} \= \frac{\epsilon}{g^2}\,
\frac{4\,x_\mu x_\nu - \eta_{\mu\nu}x{\cdot}x}{(x{\cdot}x+\delta)^3}
\ \sim\ 
\frac{\epsilon}{g^2}\,
\frac{-3\,\delta\,\eta_{\mu\nu}}{(x{\cdot}x+\delta)^3}\ .
\end{equation}

\section{Conclusion}
Starting from the geometry of the Minkowski foliations with $H^3$- (interior of the lightcone) and $\text{dS}_3$-slices (exterior of the lightcone) and exploring the origin of these symmetric spaces through cosets of the gauge group SO(1,3), we have obtained analytic solutions of Yang--Mills equation on Minkowski space that, however, diverge at the lightcone. We achieved this by first solving a SO(1,3)-invariant configuration on the cylinder $\R\times\text{SO(3)}/H$, with $H{=}\textrm{SO(3)}$ on the interior and $H{=}\textrm{SO(1,2)}$ on the exterior of the lightcone, using dimensional reduction technique of gauge theory and then translating these solutions to two different domains of Minkowski spacetime, seperated by the lightcone, with their respective foliation maps. We then computed the stress-energy tensor in both cases and found out that they have the same form. Not only this, when written compactly, we were able to cast it into a pure improvement term, a fact that helped us in finding a regularized candidate for the stress-energy tensor, defined throughout the spacetime; how this modified stress-energy tensor arise from a source term remains an open question though.

\section*{Acknowledgements}
This work was supported by the Deutscher Akademischer Austauschdienst (DAAD) with grant number 57381412. The author thanks Olaf Lechtenfeld, Gabriel Pican{\c c}o Costa and Jona R{\" o}hrig for contribution in \cite{KLPR22}. The author is also grateful to OL for a careful reading of the draft and suggesting improvements.

\end{document}